# Top-level Design and Pilot Analysis of Low-end CT Scanners Based on Linear Scanning for Developing Countries

Fenglin Liu[1,3], Hengyong Yu[2], Wenxiang Cong[3], Ge Wang[3,*]

[1]Engineering Research Center of Industrial Computed Tomography Nondestructive Testing, Ministry of Education, Chongqing University, Chongqing 400044, China
[2]Department of Biomedical Engineering, VT-WFU School of Biomedical Engineering and Sciences, Wake Forest University Health Sciences, NC, 27157, USA
[3]Biomedical Imaging Center, Center for Biotechnology and Interdisciplinary Studies, Department of Biomedical Engineering, Rensselaer Polytechnic Institute, Troy, NY 12180, USA

**Purpose:** The goal is to develop a new architecture for computed tomography (CT) which is at an ultra-low-dose for developing countries, especially in rural areas.

**Methods:** The proposed scheme is inspired by the recently developed compressive sensing and interior tomography techniques, where the data acquisition system targets a region of interest (ROI) to acquire limited and truncated data. The source and detector are translated in opposite directions for either ROI reconstruction with one or more localized linear scans or global reconstruction by combining multiple ROI reconstructions. In other words, the popular slip ring is replaced by a translation based setup, and the instrumentation cost is reduced by a relaxation of the imaging speed requirement.

**Results:** The various translational scanning modes are theoretically analyzed, and the scanning parameters are optimized. The numerical simulation results from different numbers of linear scans confirm the feasibility of the proposed scheme, and suggest two preferred low-end systems for horizontal and vertical patient positions respectively.

**Conclusion:** Ultra-low-cost x-ray CT is feasible with our proposed combination of linear scanning, compressive sensing, and interior tomography. The proposed architecture can be tailored into permanent, movable, or reconfigurable systems as desirable. Advanced image registration and spectral imaging features can be included as well.

## 1. Introduction

Computed Tomography (CT) is a mainstay in diagnostic imaging. It plays a key role in diagnosis and intervention [1]. CT system architectures have evolved in several directions. For example, most advanced systems use wide detector arrays [2], multiple sources [3], and/or very fast rotation speed, for important clinical applications (e.g., coronary artery and whole organ perfusion imaging). As a result, modern CT scanners are expensive, which are typically used by major hospitals and clinics in developed countries. On the other hand, the CT capability is hardly accessible for the patients in rural areas of developing countries, disaster scenes, and battle fields. Here we are motivated to develop an alternative CT concept that emphasizes low cost and high mobility over dose efficiency and image quality.

In the current medical CT scanners, various scanning strategies were developed, all of which are based on the rotation of an x-ray source and an associated x-ray detector array. Moreover, the filtered backprojection type algorithm [4]) remains popular in practice, which requires a series of global projections over a full-scan or half-scan angular range. As a result, a gantry with a slip ring is indispensable, which is expensive and difficult to be moved around. The gantry rotation can be as fast as 4 turns per second, which is mainly for cardiac imaging applications. To define an ultra-low-cost CT scanner, in this project we avoid challenging cardiac imaging tasks, focus on circumstances where a local x-ray CT scan is sufficient, and yet allow global reconstruction by performing a number of interior reconstructions.

The recent developed interior tomography method is an enabling technology because for the first time it offers a theoretical exact solution to the long-standing "*interior problem*" [5]. In classic CT theory, an interior region of interest (ROI) cannot be exactly reconstructed only from truncated projections. Structures outside the ROI may seriously disturb the features inside the ROI. Recently, it was demonstrated that the interior problem can be exactly and stably solved if a sub-region in the ROI is known [6-9], which is referred to as knowledge-based interior tomography. However, it is not easy to obtain precise prior knowledge of a sub-region in many cases. A further progress in development of interior tomography was inspired by compressive sensing (CS) [10-12]. Most importantly, it was demonstrated that an interior ROI can be exactly and stably reconstructed via minimizing its total variation (TV) or high order TV if the ROI is piecewise constant or polynomial [13, 14], which is referred to as CS-based interior tomography.

To address specific applications, novel CT systems and methods were investigated, which are relevant to this project. Yamato and Nakahama proposed an interesting linear scan based dental CT approach [15]. Schon et al. studied a translation-based CT data acquisition method for imaging of a cable channel inside the corner formed by two walls in a building [16]. Gao et al. developed a linear scan CT with a wide fan-angle source, a large-area detector, and an advanced image reconstruction algorithm [17]. This system shows a promise in medical imaging, industrial or security applications. However, all these studies did not use interior tomography, and were not performed for ultra-low-dose CT in general cost-effective medical imaging applications.

The above background has motivated us to design a new CT system and develop the associated CT algorithms for ultra-low-cost and yet quality performance that are instrumental to meet medical imaging needs in developing countries, especially in rural areas. In this initial investigation, we use linear scanning, fan-beam geometry, compressive sensing, and interior tomography, which can be easily extended to cone-beam geometry. The rest of this paper is organized as follows. In the next section, we will describe data acquisition and image reconstruction methods. In the third section, we present numerical simulation results to demonstrate the feasibility of the proposed system. In the fourth section, we describe two systems for horizontal and vertical patient positions respectively. In the last section, relevant issues are discussed.

## 2. Methods

### 2.A. Imaging Model

The data acquisition scheme is based on linear or translational movement of an x-ray source and a detector array. As illustrated in Fig. 1, an x-ray source and a detector array (either a linear array or a flat panel detector (FPD)) can be oppositely translated in parallel while a patient is fixed between them. In this study, we focus on 2D image reconstruction in fan beam geometry. Therefore, the detector array can be considered as one-dimensional with equally spaced bins.

When the detector array is long and there is no limit of the translational axis, the imaging geometry is illustrated in Fig. 2. The $xoy$ coordinate system is fixed with respect to the patient, with the origin at the center of an ROI of radius $r$. Given a source position $x_p$ ($p = 1, \dots, P$), where $P$ is the number of projections acquired during the translation of the source, the ray reaching a detector cell through the center of the ROI is referred to as the central ray of the projection with the central detector location $x_p$. An angle $\theta$ relative to the $x$-axis specifies a projection, an angle $\gamma$ denotes a ray within the fan beam geometry. $S_o$ is the distance from the source to the center of ROI along the $y$-axis. $S_D$ is the distance from the x-ray source to the detector along the $y$-axis. The translational movement of the source is parallel to the $x$-axis. The detector is also aligned to be parallel to the $x$-axis. In a 2D plane, each single ray can also be defined by the two parameters $t$ and $\theta$. This can be described by the following two equations:

$$\theta = \tan^{-1}\left(\frac{S_D}{x_D - x_p}\right), \tag{1}$$

where $x_D$ is the distance from the center of the ROI to a detector cell along the $x$-axis. Let $\gamma = \arctan\left(\frac{S_o}{-x_p}\right) - \theta$, we obtain the distance $t$ from the center of the ROI to the ray $(\theta, \gamma)$:

$$t = \left(x_p^2 + S_o^2\right)^{\frac{1}{2}} \sin\gamma. \tag{2}$$

The parameter $t$ in Eq. (2) is in the interval $[-r, r]$.

### 2.B. Single Translation

To reconstruct an object exactly, the classic prerequisite is a complete dataset in the parallel beam geometry; i.e., the parallel projections should be available for a 180° angular range [18], as depicted in a rectangular area between $\theta = 0$ and $\theta = \pi$ in Fig. 3. This requirement is satisfied by most CT acquisition geometries. However, the projections from a single translation (1T) of the proposed data acquisition mode only cover part of the rectangular area as shown in Fig. 3(a). This area can be determined using Eqs. (1) and (2) in the $t$–$\theta$ coordinate system. In Fig. 3(a), each line depicts the data measured at one source position, and 41 source positions were equidistantly sampled between the tube start and end positions whose geometrical parameters are in Table 1. Therefore, it is useful to determine how much information can be obtained from a single translation compared to the standard 2D parallel-beam geometry.

Table 1. Geometrical parameters for the plots in Fig. 3.

| Parameter | Value |
| --- | --- |
| Source to detector distance $S_D$ (mm) | 1000 |
| Source to object distance $S_o$ (mm) | 500 |
| Detector cell size (mm) | 1 |
| ROI diameter (mm) | 256 |
| Source translation distance $D_s$ (mm) | 800 |
| Source positions per translation $P$ | 41 |

## 2.C. Equal-angular Sampling

A single equal-spatial translation of the source not only yields insufficient data but also produces inhomogeneous data points in the sinogram as shown in Fig. 3(a). Actually, the source points for the single translation can be adjusted for better performance.

Let us define $\varphi = \theta_{(t=0)}$ as the angle between the central ray and the $x$-axis to specify a source position. Then, we have the general sampling scheme:

$$\varphi_0 = \tan^{-1}\left(\frac{S_o}{-x_0}\right)$$

$$\varphi_{P-1} = \tan^{-1}\left(\frac{S_o}{-x_{P-1}}\right)$$

$$\Delta\varphi = \frac{1}{P-1}(\varphi_{P-1} - \varphi_0)$$

$$x_n = S_o \cot(\varphi_0 + p\Delta\varphi) \tag{3}$$

where $p = 0, \cdots, P-1$, $P$ is the number of source positions per translation, $\Delta\varphi$ is an angle between two projections in the $t$-$\theta$ coordinate system (i.e., the angle between two lines in Fig. 3(b)). $\Delta\varphi$ can be directly calculated in terms of the translational endpoints of the source $x_0$ and $x_{P-1}$. As shown in Fig. 3(b), the data coverage by a single translation is significantly improved with the equal angular sampling, instead of equal spatial sampling.

## 2.D. Multiple Translations

To cover larger fraction of the full 180° angular range, the aforementioned data acquisition scheme can be applied more than once. For example, we have use either two orthogonal translations (2T) or three symmetric translations (3T). The 2T and 3T schemes are respectively shown in Fig. 5. In each translation, the source is linearly moved to acquire data on red points, and the corresponding linear detector array is marked in green. Clearly, there are gaps in 1T and 2T schemes, and overlaps in 3T scheme. Ideally, there would be neither gap nor overlap within the $t-\theta$ coordinate system but it is not practical. Nevertheless, for the 2T and 3T schemes we can optimize the translation distance $D_s$ to cover the full data range better, as shown in the 3rd column of Fig. 4.

## 2.E. Reconstruction Algorithm

The proposed imaging system can be modeled as a linear matrix equation in terms of the pixel basis:

$$AX = b, \tag{4}$$

where $b = (b^1, b^2, \cdots, b^M) \in R^M$ represents measured projection data with $M$ being the total number of the data, $X = (X_1, \cdots, X_N) \in R^N$ denotes an object to be reconstructed with $N$ being the total number of the pixels, and $A = (a_{mn})$ is the system measurement matrix with $m = 1, \cdots, M, n = 1, \cdots, N$.

While the ART is the first iterative algorithm used for CT reconstruction [19], the SART is a major refinement to the ART [20]. In recent years, acceleration techniques for iterative reconstruction were developed for SART, among which the ordered-subset (OS) scheme is a most effective one. The resultant algorithm is called OS-SART [21, 22]. Let the index set $B = \{1, \cdots, M\}$ be partitioned into $T$ nonempty disjoint subsets $B_t = \{i_1^t, \cdots, i_{M(t)}^t\}$, then

$$B = \{1, \cdots, M\} = \bigcup_{0 \leq t \leq T-1} B_t \tag{5}$$

and the OS-SART formulation is as follows:

$$X_n^{(k+1)} = X_n^{(k)} + \sum_{m \in B_{[k \bmod T]}} \frac{a_{mn}}{a_{+n}} \frac{b_m - A^m X^{(k)}}{a_{m+}} \tag{6}$$

where $a_{m+} \equiv \sum_{n=1}^N a_{mn} \neq 0, a_{+n} \equiv \sum_{m=1}^M a_{mn} \neq 0$, and $k$ is the iteration index.

The above OS-SART method can be combined with compressive sensing (CS) to improve the image quality in the case of sparse measurements. The well-known sparse transform is the discrete gradient transform (DGT). Hence, an ROI image can be reconstructed by minimizing the $\ell_1$- norm of its DGT, which is referred to as the TV minimization [13, 23] and can be expressed as

$$\min_X \|\nabla X\|_1, \text{ subject to } AX = b, X_n \geq 0, \tag{7}$$

where $\|\nabla X\|_1$ denotes TV of $X$, and

$$\|\nabla X\|_1 = \sum_{i,j} d_{i,j}, \quad d_{i,j} = \sqrt{(X_{i,j} - X_{i+1,j})^2 + (X_{i,j} - X_{i,j+1})^2} \tag{8}$$

where $X_{i,j}$ is a pixel value of a discrete 2D image, and $d_{i,j}$ is the corresponding DGT. Because a 2D image can be easily rearranged into a 1D vector, we use both $X_{i,j}$ and $X_n$ to represent the same image pixel.

The problem Eq. (7) can be solved in two loops. While an outer loop implements OS-SART to reduce data discrepancy, an inner loop minimizes the image TV. In the inner loop, a steepest gradient descent search can be used:

$$X_n^{(l+1)} = X_n^{(l)} - \lambda \omega v, \tag{9}$$

where $\lambda$ is a control parameter, $v = (\partial \|\nabla X\|_1 / \partial X_{i,j})|_{X_{i,j}=X_{i,j}[k,l]}$ is the gradient direction with respect to $X_{i,j} = X_{i,j}[k,l]$, $\omega = max(|X_n^{(l)}|)/max(|v|)$ is a scaling constant, and $k$ and $l$ are the outer and inner loop indices. The whole iterative procedure can be summarized as follows:

*Step 1.* Input data: $b$ and let $X = 0$;

*Step 2.* Calculate the current image using Eq. (6);

*Step 3.* Minimize the TV of the current image $X$ using Eq. (9);

*Step 4.* Go to *Step 2* until a stopping criterion is met.

## 3. Numerical Simulation

To evaluate the proposed data acquisition modes, we implemented a reconstruction program in MatLab for interior reconstruction in fan-beam geometry. As shown in Fig. 5, a cardiac image phantom was used for realistic simulation. This phantom consists of 700x700 pixels, each of which covers an area of 0.5x0.5 mm² resulting in a rectangular compact support of 350x350 mm². Other imaging parameters for 1T, 2T and 3T are in Table 2. The pixel size of the reconstructed image was set to 0.7x0.7 mm². In the OS-SART algorithm, the subset number was set to 6, and each subset took 40 views for each of the data acquisition modes. The pseudo-code for the TV minimization can be found [13]. The initial value of $\lambda$ was set to 0.005 and attenuated with a constant factor 0.995 after each iteration. The stopping criterion is to reach a maximum number of iterations.

Fig. 6 shows magnified 300x300 cardiac regions in the reconstructed images after 120 iterations. It can be seen that the image quality associated with 1T is inferior to that with either 2T or 3T. This is because the data acquired from 1T cannot sufficiently cover the full data range, involving truncated and limited data issues. Fig. 7 shows representative profiles. It is noticed that there exist some differences between the profiles of the original phantom and reconstructed images. This may be due to three reasons. First, interior tomography is formulated in the continuous domain which requires infinitely many views. However, we only employed 240 views in this work. Second, the iterative algorithm will converge to the true solution only after infinitely many iterations, while the iteration was stopped after 120 iterations in this study. Third, the image phantom is from a clinical application, which do not rigidly satisfy the piecewise constant image model.

In practical applications, measurement noise is unavoidable. To test the stability of the proposed data acquisition modes and the associated algorithms against data noise, we repeated the aforementioned tests from projections corrupted by Poisson noise, assuming 5x10⁴ photons per detector element [24]. The results are respectively shown in Figs. 8 and 9, which show a satisfactory stability of the imaging performance.

TABLE 2. Parameters for the numerical simulation.

| Parameter | Value |
| --- | --- |
| Source to detector distance $S_D$ (mm) | 1200 |
| Source to object distance $S_o$ (mm) | 600 |
| Detector array length (Pixel) | 400 |
| Detector pixel size (mm) | 1.0 |
| ROI diameter (mm) | 166 |
| # of source points per translation $P$ (1T) | 240 |
| # of source points per translation $P$ (2T) | 120 |
| # of source points per translation *P* (3T) | 80 |
| 1T translation distance | 1200 |
| 2T translation distance | 1200 |
| 3T translation distance | 693 |
| Reconstruction matrix | 500x500 |
| Pixel size (mm^2) | 0.7x0.7 |

## 4. System Designs

### 4.A. Vertical Architecture

Fig. 10(a) illustrates an architecture for a vertical patient position. A source-detector frame is supported on a fixed central stage. A patient is seated on a chair, which is also attached to the stage. The source and detector are respectively mounted on two motorized linear stages, and can be translated oppositely under PC control. The source-detector frame can be manually rotated around the central stage. The scanning position is selected by adjusting the heights of the source and detector panels, aided by a laser beam. During a scan, the patient remains stationary. The source-detector frame can be oriented at 2 or 3 positions to realize the 2T or 3T modes.

### 4.B. Horizontal Architecture

Fig.10(b) illustrates an architecture for a horizontal patient position. In the system, there are two source-detector pairs. A manually adjustable patient bed is for patient translation. There are four casters fixed on the frame. With 2 laser beams, the scan position can be selected, and then the casters are locked. The sources and detectors are respectively mounted on 4 motorized linear stages for the 2T scanning mode. The two source-detector pairs can be translated simultaneously.

### 4.C. System Costs

While the current low-end CT systems still cost substantially, say in a range of $300K-$600K, our proposed systems are intended to be significantly cheaper. Our system performance specification includes <1mm spatial resolution, 2 kVp settings, 3-second linear translation, comparable noise and dose indices. Given these targets, estimated system costs are listed in Table 3.

TABLE 3. Cost estimates of the proposed low-end CT scanners.

| No. | Subsystem | Supplier | Vertical System | | | Horizontal System | | |
|---|---|---|---|---|---|---|---|---|
| | | | Unit Price | Quantity | Cost | Unit Price | Quantity | Cost |
| 1 | X-ray Source | Toshiba | $5,000 | 1 | $5,000 | $5,000 | 2 | $10,000 |
| 2 | Flat Panel Detector | Samsung/ PerkinElmer | $32,000 | 1 | $32,000 | $32,000 | 2 | $64,000 |
| 3 | Mechanical Control | From China | $15,000 | 1 | $15,000 | $22,000 | 1 | $22,000 |
| 4 | Computer | | $1,500 | 1 | $1,500 | $1,500 | 1 | $1,500 |
| 5 | Software | | $2,000 | 1 | $2,000 | $2,000 | 1 | $2,000 |
| Total Cost | | | $55,500 | | | $99,500 | | |

## 5. Discussions and Conclusion

Our numerical results have shown that it is possible to image an ROI by translating an x-ray source and detector pair. Different translational modes (1T, 2T and 3T) have been evaluated with promising. While the 1T mode cannot give sufficient image quality, the optimized 2T and 3T modes seem yielding excellent imaging performance. While a simplest searching method was employed to minimize the TV, more complicated better methods can be used as well, such the soft-threshold filtering method [23]. Moreover, if the source sampling is dense enough, a filtered backprojection algorithm can be applicable in the 2T and 3T modes.

While the focus has been placed on interior reconstruction, the proposed architectures actually allow both interior and global scans. For example, a global scan can be achieved by combining multiple interior scans. A more optimized global scanning mode is to translate a detector array for each fixed source position. Multiple scanning protocols are possible, and will not be discussed further in this pilot study.

CT has been widely used for preclinical and clinical imaging since its invention. While modern CT scanners are so popular in the developed countries, low-end CT systems are on an urgent demand in developing countries, especially in the rural areas. To address this requirement, we proposed and evaluated translation based data acquisition modes and the corresponding reconstruction algorithm, which produced excellent image quality in our numerical simulation. Also, a detailed cost analysis indicates the overall system cost can be brought down to $100K or less.

In summary, we have presented two architectures for low-end CT and established their feasibility. In the next step, we plan to prototype a vertical CT system and collect real data in a small trial on large animals. Future efforts will be focused on optimization of scanning parameters, data and image registration, and cost-effective engineering/instrumentation. The main novelty is intended to define a cheapest yet most versatile low-end CT scanner, and eventually demonstrate its real-world utilities.

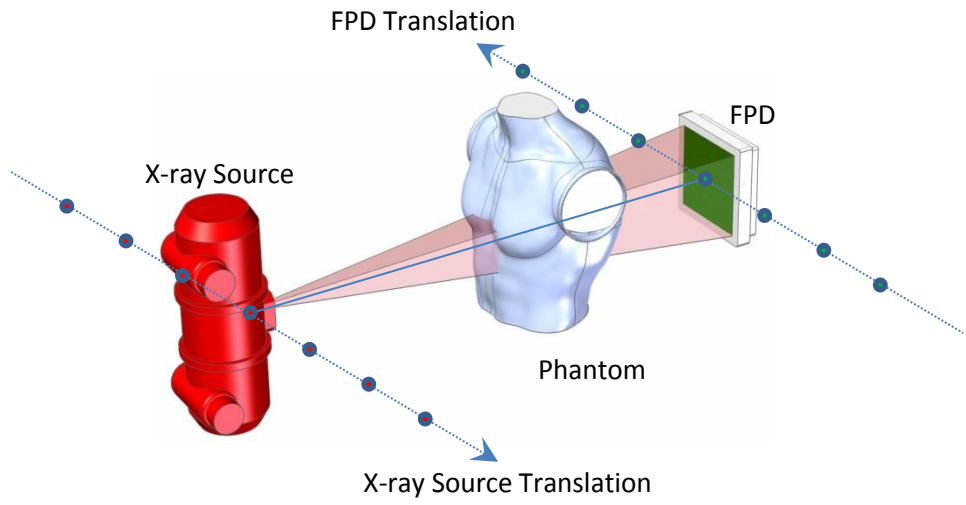

Figure 1. Linear scan based CT. An x-ray tube and a FPD are positioned face to face with a patient between them. During a scan, the source and the detector will be translated in opposite directions with the phantom stationary.

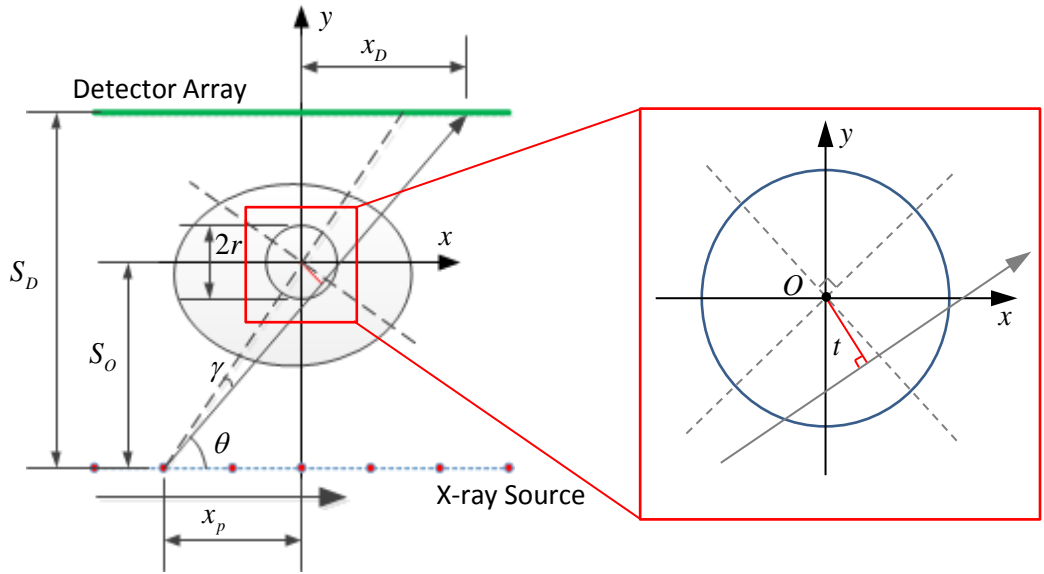

Figure 2. Imaging geometry and variables. By changing the source position, the ray measured at a fixed detector position can be changed.

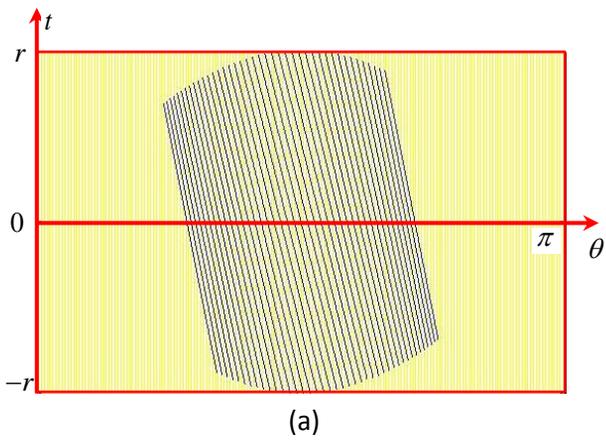 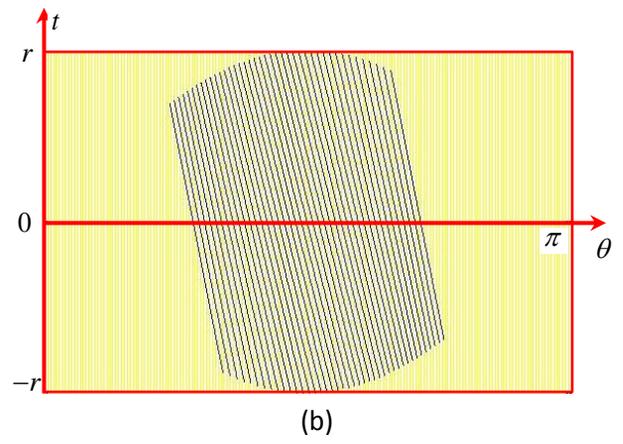

Figure 3. Rectangular data region in the *t-θ* coordinate system partially filled by a single translation of the source. (a) Sampled data from equi-spatial source positions, and (b) the counterpart data from equi-angular source positions.

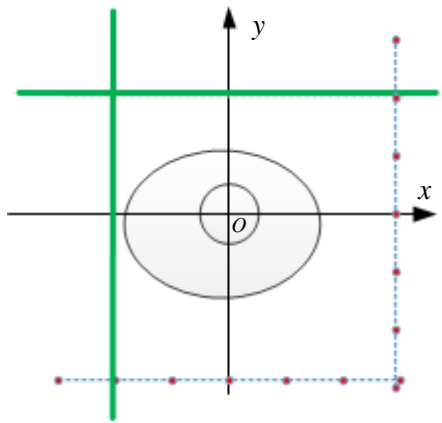
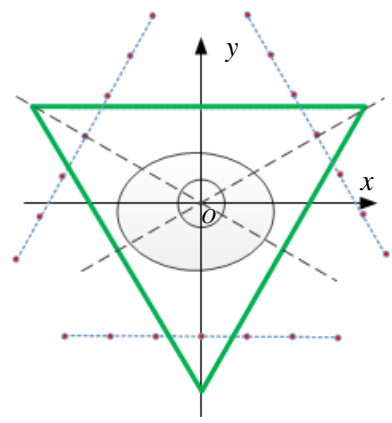
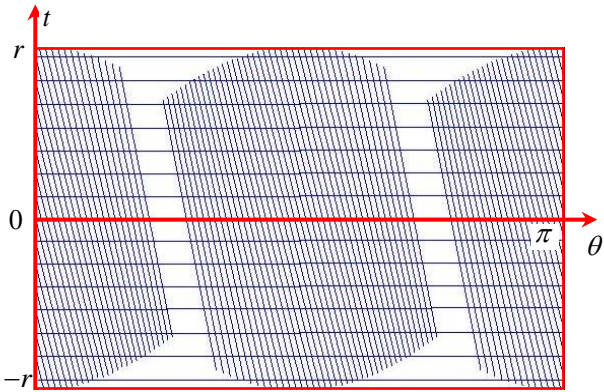
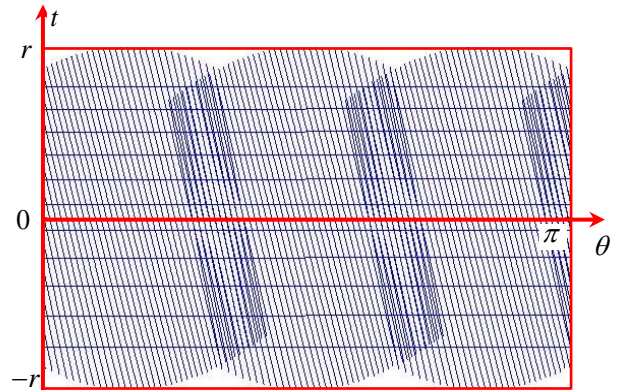
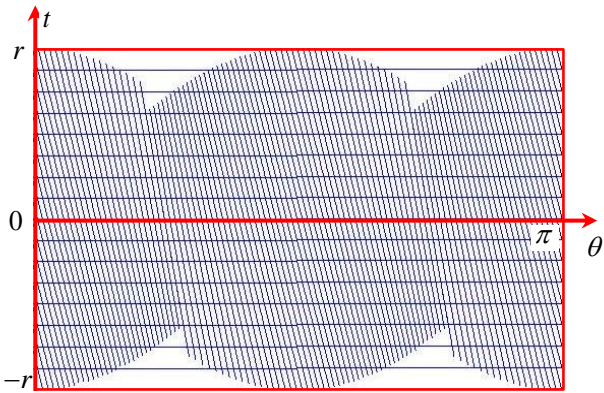
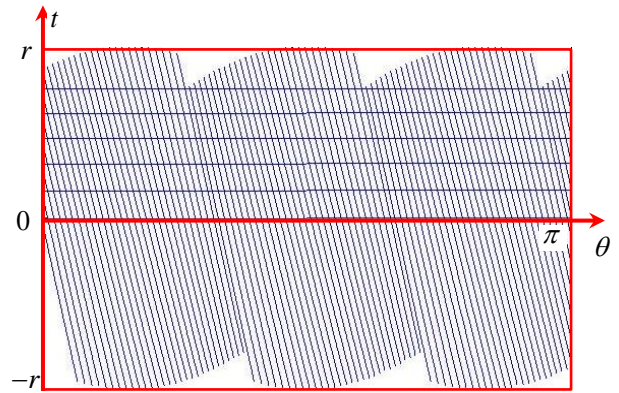

Figure 4. Different translation modes (top row) in the *t-θ* coordinate system using the equi-spatial source sampling scheme (middle row) and equi-angular counterpart (bottom row). The left and right columns are for the 2T and 3T modes, respectively.

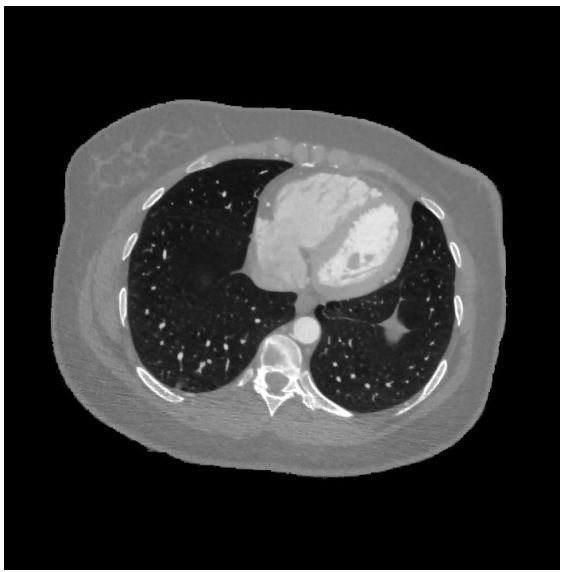

Figure 5. Cardiac image phantom in a display window of [-1000,1000]HU.

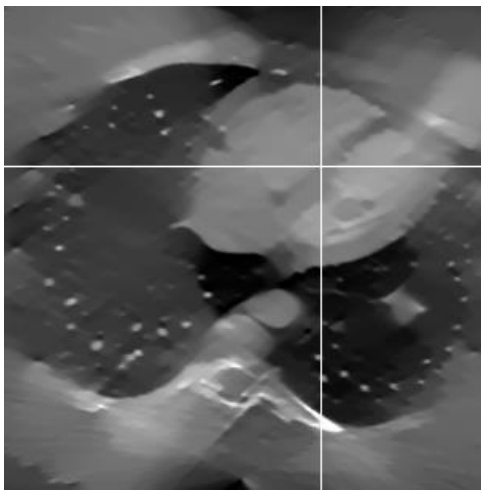 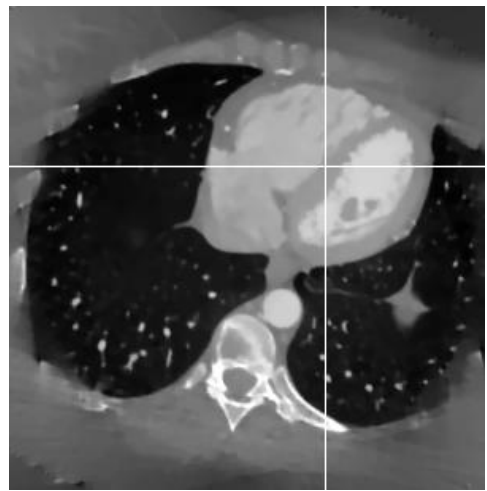

(a) (b)

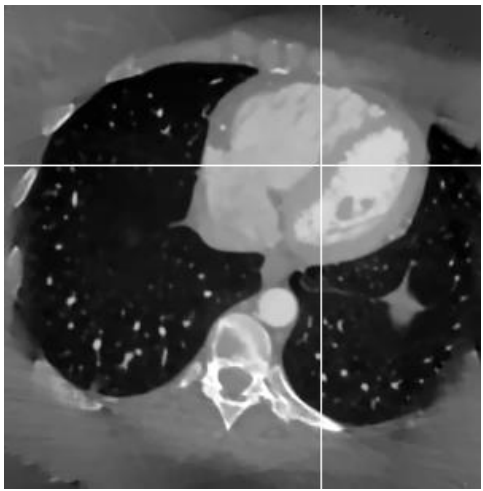 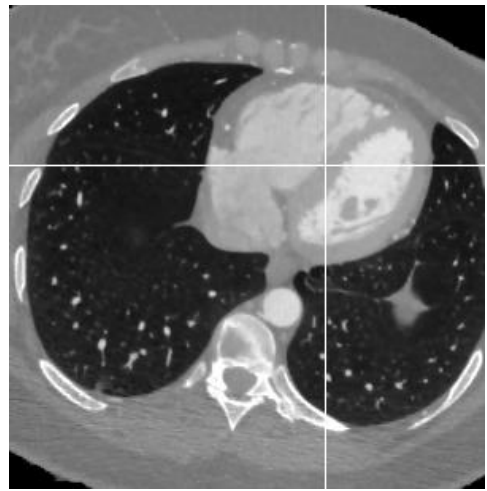

(c) (d)

Figure 6. Magnified cardiac reconstructions after 40 iterations in a display window [-1000, 1000] HU. (a), (b) and (c) are from noise-free projections acquired on in 1T, 2T and 3T scanning modes respectively, and (d) is the original phantom as a reference.

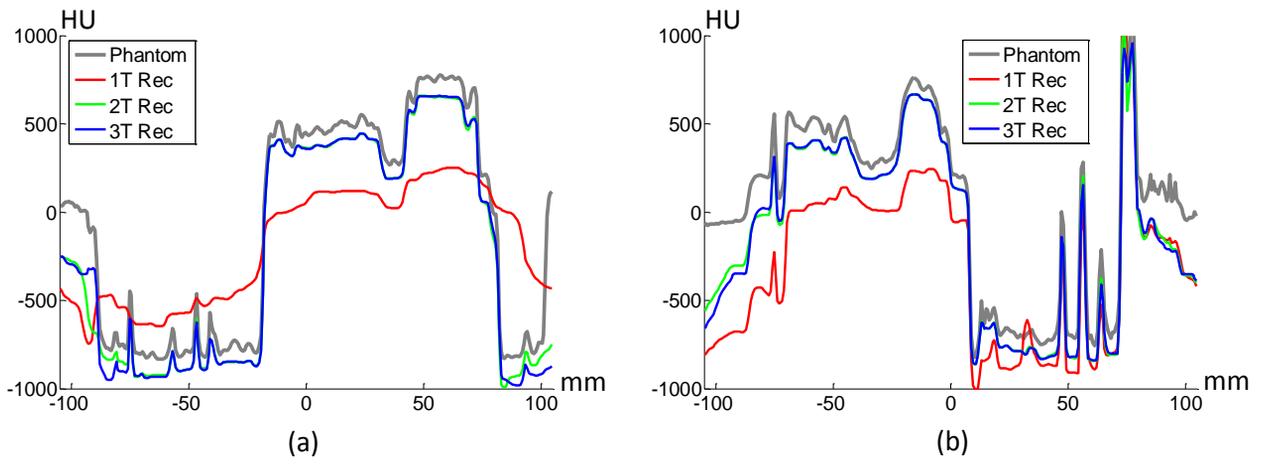

Figure 7. Representative profiles respectively along the horizontal and vertical white lines in Figure 6.

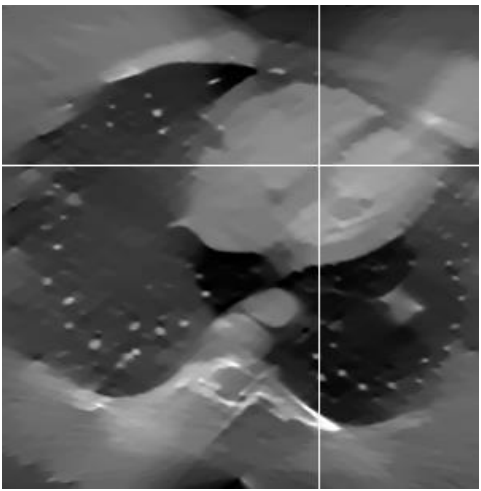
(a)

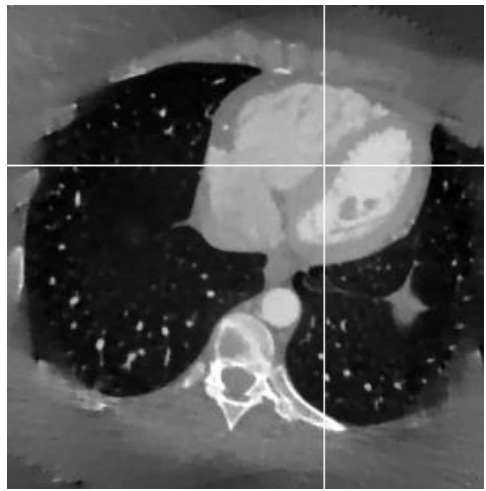
(b)

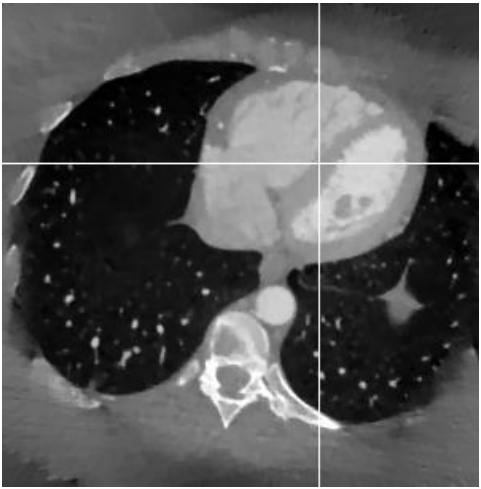
(c)

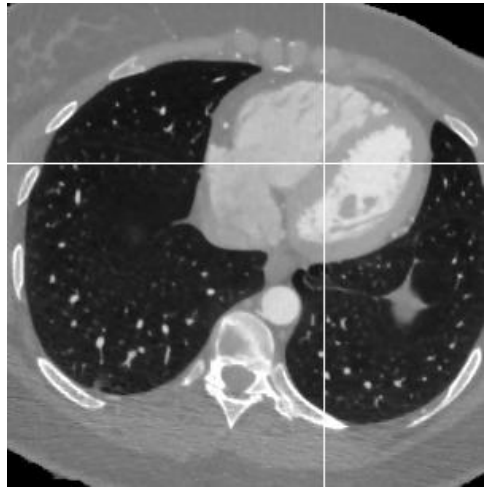
(d)

Figure 8. Same as Figure 6 but reconstructed from noisy projections.

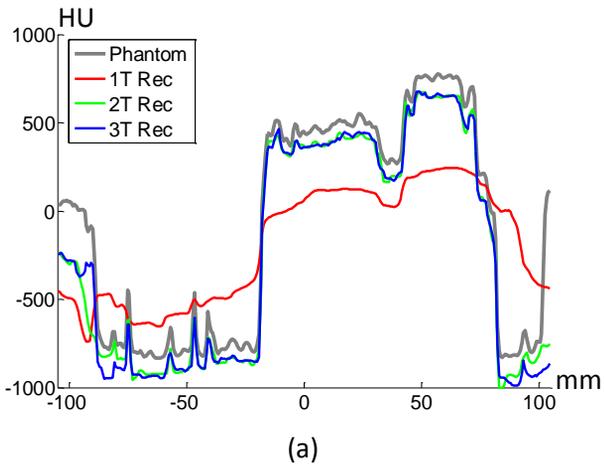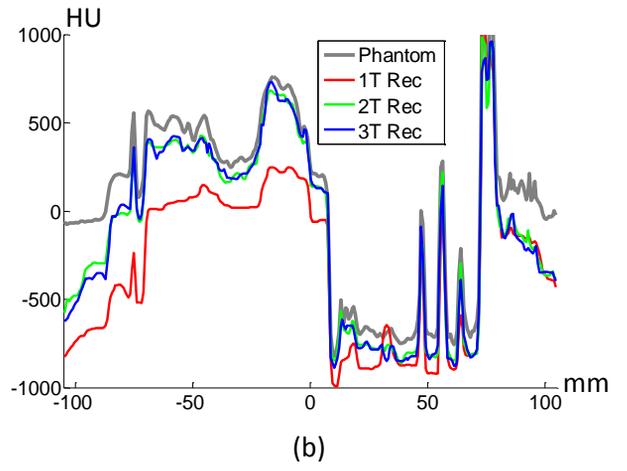

Figure 9. Same as Figure 7 but reconstructed from noisy projections.

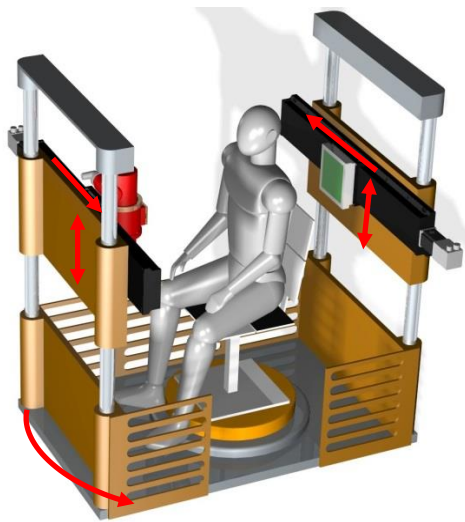 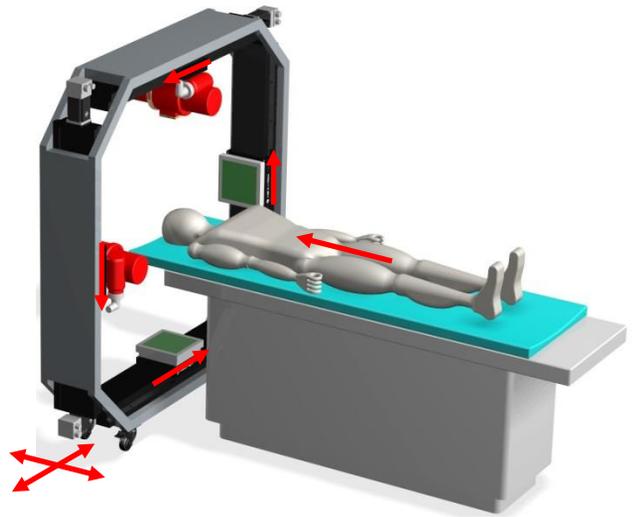

Figure 10. 3D renderings of the proposed systems. (a) A vertical version whose central stage can support a chair for a patient or allow a standing patient without the chair. The source-detector frame can be rotated to 2 or 3 orientations for 2T or 3T scans respectively; (b) a horizontal version where a patient on the bed keeps stationary during a scan, and two orthogonally arranged source-detector pairs can do a 2T scan in parallel.